\documentstyle[art12]{article}
\def\M{{\cal M}}
\def\L{{\cal L}}
\def\J{J/\psi}
\def\H{{\cal H}}
\def\J{J/\psi}
\def\U{\Upsilon}
\def\etc{\eta_c}
\def\etb{\eta_b}
\def\ee{e^+e^-}
\def\gg{\gamma\gamma}
\def\vp{\vec p}
\def\vq{\vec q}
\def\e{\varepsilon}
\begin{document}
\begin{center}

 RELATIVISTIC EFFECTS IN S-WAVE QUARKONIUM DECAYS\\[5mm]

A.P.Martynenko, V.A.Saleev\\[5mm]

Dept. of General and Theoretical Physics, Samara State University

Samara, 443011, Russia\\[10mm]

\end{center}

\begin{abstract}
The decay widths of S-wave quarkonia ($\etc,\etb\to
 \gg\mbox{ ~~and~~} \J,\U\to\ee$) are calculated on the basis of
a quasipotential approach. The nontrivial dependence on relative
quark motion of decay amplitude is taken into consideration via
quarkonium wave function. It is shown that relativistic corrections
may be large (10-50 \%) and comparable with QCD corrections.
\end{abstract}

\section{Introduction}
The constituent quark models \cite{1} are being not well grounded
from the point of view of quantum field theory, as the QCD sum rules
method \cite{2} or lattice QCD approach \cite{3}, but they give us
 a opportunity
to describe most of all existing experimental data. The great success in
description of heavy meson static characteristics and decay widths
has been obtained on the basis of a relativistic quark models \cite{4,5}.
As it was shown, relativistic corrections are not small for mesons
containing comparably light charm quark \cite{5a}. In this paper we
present the results of calculation of the relativistic corrections
in S-wave quarkonium decays
($\etc,\etb\to\gg\mbox{~and~} \J,\U\to\ee$) using the approach
based on a Logunov-Tavkhelidse local quasipotential equation \cite{6}
and perturbative QCD methods.

\section{Decay amplitude in quasipotential approach}

The meson decay amplitude may be
presented as the product of an parton decay amplitude $\M_H(q_1,q_2)$
with amputate quark lines
times a Bethe-Salpeter wave function $\Psi(q_1,q_2)$:
\begin{equation}
\M=\int\frac{d^4q_1}{(2\pi)^4}Sp\left[\Psi(q_1,q_2)\M_H(q_1,q_2)\right],
\end{equation}
where $\Psi(q_1,q_2)$ describes two-quark bound state,
 $\M_H(q_1,q_2)$ describes  transition between initial two-quark state
and final state of free particles, $Sp$ means the trace over spin and
colour indexes. The wave function $\Psi(q_1,q_2)$  satisfies two-particle
Bethe-Salpeter equation, which has not physically interesting exact
solutions. That is why the redefinition of the wave function for calculation
of decay amplitude $\M$ is needed. We must to introduce new wave function,
which has clear physical interpretation and it is determined by more
simple bound state equation.

Let us define vertex function $\Gamma(q_1,q_2)$
as follows:

\begin{equation}
\Psi(q_1,q_2)=\frac{\hat q_1+m}{q_1^2-m^2+io}\Gamma(q_1,q_2)
\frac{\hat q_2-m}{q_2^2-m^2+io}.
\end{equation}

Transforming (1) to the center-of-mass reference frame and using projection
operator decomposition for quark propagator on positive and negative
energy states
\begin{equation}
 \frac{\hat p+m}{p^2-m^2+io}=\frac{1}{2\e(\vp)}
 \left[\frac{U^{\alpha}(\vp)\bar U^{\alpha}(\vp)}{p^o-\e(\vp)+io}+
 \frac{V^{\alpha}(-\vp)\bar V^{\alpha}(-\vp)}{p^o+\e(\vp)-io}\right],
\end{equation}
we can write amplitude $\M$ in following form:
\begin{eqnarray}
 \M&=&\frac{1}{(2\pi)^3}\int\frac{d^3q}{2\e(\vq)}\int\frac{dq^o}{2\pi}
 \frac{(M-2\e(\vq))}{(q^o-\e(\vq)+io)(q^o-M+\e(\vq)-io)}\nonumber\\
&& Sp\left[\Psi_M(\vq)\hat R_s(\vq)\M_H(q_1,q_2)\right],
\end{eqnarray}
where $\e(\vp)=\sqrt{\vp^2+m^2}$, $q_1=(q^o,\vq),q_2=(M-q^o,-\vq)$,
$M$ is the mass of bound state,
$\Psi_M(\vq)$ is single time bound state wave function which was obtained
  via projection the Bethe-Salpeter amplitude on positive energy
  state \cite{7}:
 \begin{equation}
 \bar U(\vq)\Gamma(q_1,q_2)V(\vq)=2\e(\vq)(M-2\e(\vq))\Psi_M(\vq).
 \end{equation}

Note, that for S-wave bound states which are discussed here,
 $\Psi_M(\vq)=\Psi_M(|\vq|)$.

The spin property of the two-quark bound state is described by
relativistic projection operator, in which the exact dependence on
vector momentum $\vq$ must be considered.
We write it as in \cite{8}:

\begin{equation}
 \hat R_s(\vq)=\frac{(\hat q+m)(1+\gamma_o)\hat s(m-\hat q')}
  {2\sqrt{2}(\e(\vq)+m)},
\end{equation}
where $\hat s=\hat e=e_{\mu}\gamma^{\mu}$ for vector $^3S_1$ state,
$e_{\mu}$ is polarization vector, and $\hat s=\gamma_5$ for pseudoscalar
$^1S_o$ state, $q=(\e(\vq),\vq),\quad q'=(\e(\vq),-\vq)$.

The relative energy $q^0$ integral in (4) is defined by poles
of integrated function. The pole $q^o=\e(\vq)-i0$ in lower halfplane of
complex variable $q^0$ gives main contribution. It is obviously that
nontrivial contribution of other poles in parton amplitude $\M_H(q_1,q_2)$
is suppressed by factor
$\Delta\approx (M-2\e(\vq))/m$,
which however is not very small in the case of charmonium state decays
and it  will be discussed separately.

The nonrelativistic approximation corresponds to limit $\vq\to0$.
In this case the meson decay amplitude is equal to
\begin{equation}
\M_{0}=\frac{\Psi_M(0)}{\sqrt{2M}}Sp\left[\hat R_{s0}\M_H(0,0)\right],
\end{equation}
where
$$\Psi_M(0)=\int\frac{d^3q}{(2\pi)^3}\Psi_M(\vq),\mbox{ and  }
 \hat R_{s0}=m(1+\gamma_0)\hat s/\sqrt{2}.$$
The expression (7) is agree with usual used formula in potential
quark models \cite{1}.

The wave function $\Psi_M(\vp)$ of two-particle bound system with mass $M$
satisfies to quasipotential equation in the center-of-mass reference
frame:
\begin{equation}
(M-2\e(\vp))\Psi_M(\vp)=\int\frac{d^3q}{(2\pi)^3}V(\vp,\vq,M)\Psi_M(\vq),
\end{equation}
where $V(\vp,\vq,M)$ is so-called quasipotential, which is calculated
using two-particle scattering off-shell amplitude,
$\vp$ is a relative momentum of particles: $\vp=(\vp_1-\vp_2)/2$.
For more comfortable investigation bound state wave function and
mass spectrum,  we transform eq. (8) to a local form \cite{9}:
 \begin{equation}
 \left(\frac{b^2(M)}{2\mu_R}-\frac{\vp^2}{2\mu_R}\right)\Psi_M(\vp)=
 \int\frac{d^3q}{(2\pi)^3}V(\vp,\vq,M)\Psi_M(\vq),
 \end{equation}
where $b^2(M)=(M^2-4m^2)/4$ is the square of relative momentum of
particles on energy shell  $(M=2\e(\vp))$,  $\mu_R=M/4$ is relativistic
reduced mass.
Than we obtain from eq.(9) the bound state equation in coordinate space:
\begin{equation}
\left(-\frac{\Delta^2}{2\mu_R}+U(\vec r)\right)\Psi_M(\vec r)=
 E_R\Psi_M(\vec r),
\end{equation}
 where
 $$E_R=\frac{b_R^2}{2\mu_R}=\frac{M^2-4m^2}{2M}.$$

In the case of spherical symmetry potential  $U=U(r)$
 we have for radial part of wave function:
\begin{equation}
\frac{d^2\chi}{dr^2}-2\mu_R\left(U(r)-E_R\right)\chi=0,
\end{equation}
where
 $$\chi(r)=\sqrt{4\pi}r\Psi_M(r).$$

   Taking into account the accuracy of our calculations, we shall use
non-relativistic approximation of quark-antiquark interaction potential.
As usual we present it using a composition of Coulomb and linear components:
\begin{equation}
 U(r)=-\frac{b}{r}+ar+c.
\end{equation}

In our calculation we use following numerical value for parameters of
interactive potential:
$b=0.25,\quad a=0.27 \mbox{ GeV}^2,
\quad c=-0.76 \mbox{ GeV}$ - for  $c-$quarks;
 $b=0.4,\quad a=0.25 \mbox{ GeV}^2,
\quad c=-0.3 \mbox{ GeV}$ - for $b-$quarks \cite{1}.
 Because we don't take into consideration spin dependence of quark-antiquark
 potential, it is necessary to use quark masses as free parameters
 for right description of mass spectra of quarkonium  states. We obtained:
 $m_c=1.62$ GeV
 and $m_b=4.87$ çÜ÷ for $^3S_1$ state, $m_c=1.55$ GeV É $m_b=4.84$
 GeV for $^1S_0$ state. In this case we have: $M_{\J}=3.10$ GeV,
$M_{\etc}=2.98$ GeV, $M_{\U}=9.45$ GeV, $M_{\etb}=9.4$ GeV.

So, solving numerically bound state equation (11) with potential (12),
we obtain quarkonium wave function in coordinate space. After
numerical Fourier transformation we obtained the wave function in
momentum space,
which was used for estimation of relativistic effects in S-wave
quarkonium decays.

\section{Two-lepton decays $J/\psi,\Upsilon\to \ee$}

 Two-lepton decay of $^3S_1$  state of heavy quark-antiquark
 bound system ( for example $J/\psi$) is very interesting both
 for theory and
 experiment in particle physics. First, two-lepton mode of
 $\J$ decay is good trigger in experiment, where $\J$ production is
studied. Second, as it is usually expected, we can used in first
approximation simple non-relativistic description for bound state
of heavy quarks and study relativistic effects as small corrections.
 From the point of view perturbative QCD and motivation of quark-antiquark
potential, the small value of running constant $\alpha_s(m_c^2)=0.3$
 is also very important. That is why calculation $\J$ or $\U$ decay
 widths is a good test of our knowledge about quark-antiquark
 interaction at large distance as well as about role of relativism
 in description of heavy quark bound states.

The amplitude of decays $\J,\U\to\ee$ has the form:
\begin{eqnarray}
\M(J/\psi\to\ee)&=&\sqrt{2M}\int\frac{d^3q}{2\e(\vq)(2\pi)^3}
\int\frac{dq^o}{2\pi}Sp\left[R_J(\vq)\M_H(q_1,q_2)\right]\nonumber\\
&&F_c\frac{(M-2\e(\vq))\Psi_M(q)}{(q^o-\e(\vq)+io)(q^o-M+\e(\vq)-io)},
\end{eqnarray}
where $q_1=(q^o,\vq),\quad q_2=(q^o-M,-\vq)$, $F_c=\sqrt{3}$ is colour
factor. The relativistic normalization for quasipotential wave
function gives the multiplier $\sqrt{2M}$  in (16). The "hard" part of
amplitude can be written as:

$$\M_H(q_1,q_2)=\frac{e^2e_c}{M^2}\gamma^{\mu}\bar U_e(k_1)\gamma_{\mu}
  V_e(k_2).$$

In the case of $\J,\U\to\ee$ decays, amplitude $\M_H$ doesn't depend on
relative quark energy and the $q^o$ integration is carried out simply.
After sum over lepton polarizations, the squared modulus of amplitude
takes the form
\begin{eqnarray}
&&|\M(J/\psi\to\ee)|^2=\frac{e^4e_c^2}{M^3}\int\frac{d^3q}{(2\pi)^3}
 \frac{\Psi_M(\vq)}{\e(\vq)(\e(\vq)+m)}\nonumber\\
 && \int\frac{d^3p}{(2\pi)^3}\frac{\Psi_M(\vp)}{\e(\vp)(\e(\vp)+m)}
 \L_{\mu\nu}(k_1,k_2)\H^{\mu}(\vq)\H^{\nu}(\vp),
\end{eqnarray}
where
$$\L_{\mu\nu}(k_1,k_2)=4\left[k_{1\mu}k_{2\nu}+k_{1\nu}k_{2\mu}-
 (k_1k_2)g_{\mu\nu}\right],$$
$$\H^{\mu}(\vq)=Sp\left[(\hat q_1+m)(1+\gamma_o)\hat e_J(\hat q_2-m)
   \gamma^{\mu}\right].$$

Using the condition for $\J$ polarization vector
$$\sum_{s_J}\e_{\mu}\e_{\nu}=-g_{\mu\nu}+P_{\mu}P_{\nu}/M^2,$$
we average (14) over $\J$ polarizations and than carry out the angle
integration. Finally we obtain:
\begin{eqnarray}
\frac{1}{3}\sum_{s_J}|\M|^2&=&\frac{16\alpha^2e_c^2}{9\pi^3M}
(m^2I_0+3m^3I_1+2I_2)^2,
\end{eqnarray}
where
\begin{equation}
I_0=\int\frac{q\chi(q)dq}{\e(\vq)(\e(\vq)+m)},
I_1=\int\frac{q\chi(q)dq}{(\e(\vq)+m)},
I_2=\int\frac{\e(\vq)q\chi(q)dq}{(\e(\vq)+m)}.
\end{equation}
The width of  $\J\to\ee$ decay and amplitude are connected as follows:
\begin{equation}
\Gamma(\J\to\ee)=\frac{1}{16\pi M}\frac{1}{3}\sum_{s_J}|\M|^2.
\end{equation}

In non-relativistic approach it reads:
\begin{equation}
\Gamma_0(\J\to\ee)=16\pi\e_c^2\alpha^2\frac{|\Psi_M(0)|^2}{M^2}.
\end{equation}

Our numerical calculation gives for $\J$ particle
$\Gamma/\Gamma_0=0.89$ and for $\U$ particle - $\Gamma/\Gamma_0=0.97$.
So, the values of relativistic corrections in two-lepton decays of
$\J,\U$ particles are not small, but they are less than QCD corrections
  in next order of $\alpha_s$ \cite{10}:
\begin{equation}
\Gamma_{QCD}=\Gamma_o\left(1-\frac{16}{3}\frac{\alpha_s}{\pi}\right).
\end{equation}
Putting up   $\alpha_s(m_c^2)=0.3$ and $\alpha_s(m_b^2)=0.2$, we obtain
$\Gamma_{QCD}/\Gamma_o(\J)\approx 0.5$ and
  $\Gamma_{QCD}/\Gamma_o(\U)\approx 0.7$. Note, that our calculations give
 more large relativistic corrections than estimation in binding energy $\e$
 approximation \cite{5}:
\begin{equation}
\Gamma=\Gamma_o\left(1-\frac{1}{3}\frac{\e}{m}\right),
\end{equation}
This equation gives $\Gamma/\Gamma_0=0.96$ for $\J$ and
$\Gamma/\Gamma_0=0.98$ for $\U$ particles.
The difference may be explained using following fact. We exactly take into
consideration relative quark motion in amplitude and in wave function,
but in \cite{5} $\vp^2/m^2$ approximation in amplitude is used and
wave function is presented only in origin.

\section{Two-photon decays $\eta_c,\eta_b \to \gg$}

The two-photon decays of $\etc$ and $\etb$ particles are not studied
sufficiently in experiment opposite to lepton decays of $\J$
and $\U$ particles. The experimental uncertainty of decay width
is equal to 50\% for $\etc\to\gg$ decay and data is absent for
$\etb\to\gg$ decay \cite{11}. Contradict to this fact, these
decays are very interesting for discrimination of many models
which describe heavy quark bound state and QCD properties
at large distance \cite{12}.

The decay amplitude $\etc\to\gg~~(\etb\to\gg)$ may be presented
in the form:
\begin{eqnarray}
\M(\eta_c\to\gg)&=&\sqrt{2M}\int\frac{d^3q}{2\e(\vq)(2\pi)^3}
\int\frac{dq^o}{2\pi}Sp\left[R_{\eta}(\vq)\M_H(q_1,q_2)\right]\nonumber\\
&&F_c\frac{(M-2\e(\vq))\Psi_M(\vq)}{(q^o-\e(\vq)+io)(q^o-M+\e(\vq)-io)}.
\end{eqnarray}
The "hard" part of amplitude $\M(\eta_c\to\gg)$ may be written
as follows:
\begin{eqnarray}
\M_H(q_1,q_2)&=&e^2e_c^2 e_{\nu}(k_1)e_{\mu}(k_2)
[\gamma^{\nu}\frac{\hat q_1-\hat k_1+m}{(q_1-k_1)^2-m^2}\gamma^{\mu}
 +\nonumber\\
&&\gamma^{\mu}\frac{\hat q_1-\hat k_2+m}{(q_1-k_2)^2-m^2}\gamma^{\nu}].
\end{eqnarray}

The integral function in (32) has two poles in plane of complex variable $q^o$:
$q_1^o=\e(\vq)-io$ or
$q_1^o=M-\e(\vq)+io$ as well as  $q_1^o=k_1^o\pm\sqrt{k_1^{o2}+m^2}\mp io$
 and $q_1^o=k_2^o\pm\sqrt{k_2^{o2}+m^2}\mp io$ .

The contribution of other poles from "hard" part of amplitude is
suppressed comparably the contribution of pole $q_1^0=\e(\vq)$ by
factor $\Delta\approx 0.1$ for $\etc$ and $\Delta\approx 0.03$ for
$\etb$ particles. We will discuss it below.

After integration equation (32) over $q_1^0$, the contribution of the
pole $q_1^0=\e(\vq)-io$ reads:
\begin{eqnarray}
 \M(\eta_c\to\gg)=\sqrt{2M}\int\frac{d^3q}{(2\pi)^3}\frac{\Psi_M(\vq)}
 {2\e(\vq)}F_c
Sp\left[\hat R_{\eta}(\vq)\M_H(\vq)\right],
\end{eqnarray}

\begin{equation}
 |\M|^2=32F_c^2e_c^4(4\pi\alpha)^2\frac{MJ^2}{4m^2},
\end{equation}
where
$$J=\frac{m^2}{(2\pi)^3}\int\frac{d^3q}{2\e(\vq)}\Psi_M(\vq)\frac{1}{q}
  \ln\left[{\frac{\e(\vq)+q}{\e(\vq)-q}}\right]$$

In non-relativistic approach we obtain  well known result from (32):
\begin{equation}
 |\M_0|^2=32F_c^2e_c^4(4\pi\alpha)^2\frac{|\Psi_M(0)|^2}{M}.
\end{equation}

After numerical calculation of (35) we obtained $\Gamma/\Gamma_0=0.53$
for $\etc\to\gg$ decay and $\Gamma/\Gamma_0=0.71$ for $\etb\to\gg$ decay.
The increase of the relativistic corrections in two-photon decays as
compare with two-lepton decay is explained by the nontrivial dependence
of the two-photon decay amplitude from relative quark momentum $\vq$.
Obtained value of relativistic correction for two-photon decay width
is approximately equal to QCD correction \cite{10}:

  $$\Gamma_{QCD}=\Gamma_o\left(1-\frac{\alpha_s}{\pi}(\frac{20}{3}
     -\frac{\pi^2}{3})\right).$$

For estimation of contributions from the other poles in "hard" part of decay
amplitude, we keep in mind that we can to ignore the dependence of
amplitude on the vector momentum $\vq$. Thus the given contribution
elementary related with non-relativistic approximation:
 \begin{equation}
\M_2=\M_0(1+\delta),
 \end{equation}
where
$$\delta=\frac{M-2m}{\sqrt{8}m}.$$

For $\etc\to\gg$ decay we obtained: $\delta=-0.027$. Note, that the
sign of this correction is strongly dependent on choice of quark-antiquark
potential.

\section{Conclusions}

Using relativistic quasipotential approach, we have calculated decay widths
for S-wave quarkonia: $\eta_c,\eta_b\to\gamma\gamma,\quad J/\psi,\Upsilon
\to e^+e^-$. Nontrivial dependence on relative quark motion of
decay amplitude is taken into account both in parton amplitude as in
quarkonium wave function.

Our conclusions are:
\begin{enumerate}
\item Relativistic corrections for S-wave quarkonium decay are not
 small (10-50\%) and comparable with QCD corrections.
\item Quasipotential approach gives more large value for relativistic
 corrections than binding energy approximation.
\item Additional poles in a relative-energy integral determine non vanishing
 contribution in decay width of quarkonia.
\end{enumerate}

\section{Aknowlegements}
We are grateful to R.Faustov and V.Galkin for the stimulating
discussion of the bound state problem and C.Carimalo for critical
remarks and discussion. This work is supported in part
by Russian Foundation of Basic Research under Grant 93-02-3545 and
Russian State Committee on High Education under Grant 94-6.7-2015.

\end{document}